# Holographic Maxwellian near-eye display with adjustable and continuous eye-box replication

SHIJIE ZHANG, JUAN LIU, ZHIQI ZHANG

*Beijing Engineering Research Center for Mixed Reality and Advanced Display, School of Optics and Photonics, Beijing Institute of Technology, Beijing 100081, China*
*Corresponding author:juanliu@bit.edu.cn*

**The Maxwellian display presents always-focused images to the viewer, alleviating the vergence-accommodation conflict (VAC) in near-eye displays (NEDs). However, the limited eyebox of the typical Maxwellian display prevents it from wider applications. We propose a holographic Maxwellian near-eye display with adjustable and continuous eye-box replication. Holographic display provides a way to match the human pupil size with the interval of the replicated eyeboxes, making it possible to eliminate or alleviate double image or blind area problem, which exists long in eyebox expansion for Maxwellian display. Besides, seamless image conversion between viewing points has been achieved through hologram pre-processing. Optical experiment confirms that the interval between replicated eyeboxes is dynamically adjustable, ranging from 2mm to 6mm. The proposed display can present always-focused images and seamless conversion among viewpoints with 5.32° horizontal field of view, and 9mmH × 3mmV eyebox.**

Augmented reality (AR) technology has attracted ever-increasing intention in the past decades. It could be applied in various fields such as military, surgery, education, entertainment, communication and provide brand new ways how people interactive with the real world [1, 2]. Among AR devices, compact and portable near-eye displays (NEDs) hold the most potential for development. For NEDs, one of the most issues that hinders wider commercial applications is the mismatch between users' vergence and accommodation which will cause discomfort and eye fatigue [3, 4]. Three-dimensional (3D) display techniques are good choice to solve this problem. NED systems based on light field display [5], multi-focal plane display [6], holographic display [7-9] and other 3D techniques have been studied and proposed. Beyond that, Maxwellian display, also known as retinal projection display, is also a perfect solution to the vergence-accommodation conflict through converging the light rays to the eye pupil. The images are projected directly onto the retina so the viewers could perceive always-in-focus images regardless of the focal length of crystalline lens. With the effective eye pupil turning into a convergent point, the depth of focus of Maxwellian display becomes extremely large but meanwhile, the eyebox of the system is narrowed and fixed around the single focus point. The images may be lost when there is a device displacement or eyebox rotation.

To expand the eyebox of Maxwellian display system, there are generally two strategies: eyeball tracking and eyebox replication. Eyeball tracking systems are equipped with additional eyeball-tracking devices which are synchronized with the display module. Takaki et al[10] proposed a holographic Maxwellian display with flexible retinal image generation. A spatial light modulator (SLM) is employed to generate convergent point and provide eyebox movement electrically along with eyeball movement. This kind of system may be bulky in size and the performance relies on more precise and lower-latency eyeball-tracking systems. The eyebox replication scheme aims to replicate the eyebox of the system in one-dimension or two- dimension so that there is always at least one convergent point lying in the pupil range when the eye moves. In this way, the viewer could see continuous image with an enlarged visual range. Eyebox replication can be realized via spatial multiplexing, temporal multiplexing, new materials, etc. Kim et al[11] designed an eyebox replication Maxwellian display system using a multiplexed holographic optical element (HOE). Multiple concave mirrors are recorded into a single HOE. The eyebox is replicated in horizontal direction with only 3 points and the eyebox range is enlarged to 9mm×3mm. Lin et al[12] utilized a thin-film beam deflector to realize two-dimensional eyebox replication of Maxwellian NED. The beam deflector is composed of two Pancharatnam-Berry deflectors (PBDs) and a quarter-wave plate (QWP). Each PBD could deflect light in one dimension and the QWP is used to convert the polarization after the first PBD so that the light is deflected two-dimensional after the whole film. A single focus spot is multiplexed into 3×3 and the eyebox is enlarged to 9mm×9mm. Eyebox replication could enlarge the eyebox of the system effectively. Nonetheless, nearly all the Maxwellian NEDs with eyebox replication provides settled replication intervals. Shi et al[13] proposed a Maxwellian see-through NED based on a multiplexed holographic optical element (HOE) and polarization gratings (PGs) to extend the eyebox by tunable viewpoint multiplication. Nevertheless, the adjustment ability of the viewpoint is imited and image shift or tilt is still inevitable, which limits its practicability. It is known that the pupil size would change with a range of about 2mm to 6mm when human eyes receive different light intensity. In this case, if the interval between replicated eyebox is smaller than the pupil size of human eye, the viewer will see two images simultaneously when the eyeball moves from one convergent point to adjacent point of an eyebox replication Maxwellian NED. If the interval is larger than the pupil size, there will be an information loss between adjacent eyebox. Besides, there are always obvious image shifts when moving among the different viewpoints.

In this Letter, we proposed a holographic Maxwellian near-eye display system with adjustable and continuous replication eyebox. To match the variable pupil size of human eye, the proposed system can flexibly encode complex wavefronts into a phase-only hologram (PO-CGH), so that the interval between replicated eyebox is dynamically adjustable. The images are also continuous between the neighboring viewpoints by a simple pre-compensation method.

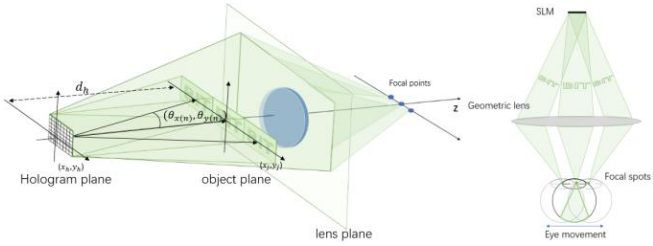

Fig. 1. Schematic diagram of eyebox expanded holographic system.

The basic holographic configuration to expand eyebox can be illustrated in Fig. 1. fig1. The model consists of a hologram plane, a reconstruction plane, convergent lens and a pupil plane. In order to fulfill an eyebox-expanded Maxwellian view, we compute a phase-only hologram that combines several objects corresponding to different viewing spots, so that the reconstructed images can converge at pupil plane behind convergent lens and form a spots array.

The complex-amplitude distribution of different convergent spots at the hologram plane can be calculated using the point-based method as:

$$H_{holo}(x_h, y_h) = \sum_{j=1}^{N} A_j \cdot \exp\left[ik\sqrt{(x_h - x_j)^2 + (y_h - y_j)^2 + (d_h - z_j)^2}\right]$$

where $(x_h, y_h)$ is the coordinates on hologram plane, N is the number of object points and λ is wavelength, k is wave number, and $A_j$ are the object point J's coordinates and amplitude, and $d_h$ is the distance between the 3D object and hologram plane. Applying an individual plane carrier wave at angle $(x_h\theta_{x(n)}, y_h\theta_{y(n)})$ to the hologram, it can be written as:

$$H_{holo(n)}'(x_h, y_h) = H_{holo}(x_h, y_h) \cdot \exp\left[\frac{ik(x_h\theta_{x(n)} + y_h\theta_{y(n)})}{d_h}\right]$$

Here n denotes the sub-hologram index and the n-th sub-hologram can be written as:

$$H_{holo(n)}'(x_h, y_h) = a(x_h, y_h) \cdot \exp\left[i\varphi_{(n)}(x_h, y_h)\right]$$

Then we can get the PO-CGH:

$$P(x_h, y_h) = \varphi_{(n)}(x_h, y_h)$$

To expand the eyebox, we further calculate a hologram to generate a viewpoint array. To get the shifted viewpoint, in this point-based method, different plane carriers can be added to the hologram to steer the beam toward the wanted directions. After getting the holograms corresponding to different viewpoints, we could multiplex all sub-holograms into a composite hologram or refresh the set of holograms to generate a pupil array.

Expanding eyebox through pupil duplication has been reported for a long time. But the interval of the pupils in these proposed theories are usually fixed. It is known that the pupil size would change with a range of from 2mm to 8mm, when human eyes receive different light intensity. When there is a mismatch between the pupil size and viewpoints interval, the image would disappear or overlap, which greatly reduces the practicality of Maxwellian display. Nevertheless, this shortcoming can be avoided by adjusting the interval to match the human pupil diameter.

The method to adjust pupils' interval can be illustrated in fig 2 (c). The positions of different viewpoints are corresponded to the shifted angle of the object and the focal length $f'$ of convergent lens. The relation between them can be written as:

$$H = f' \tan\theta$$

Here H stands for the interval between the viewpoints. The interval can be adjusted by choosing proper $\theta$ according to the human pupil diameter.

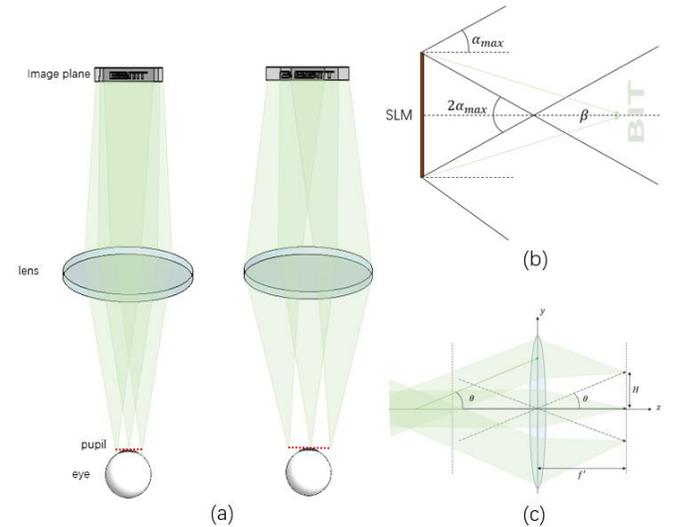

Fig. 2. The analysis of pupils' interval adjustment. (a) Schematic diagram of interval adjustment. (b) Diffractive angle reconstruction on the adjustable range of the pupils' interval. (c) The relation between focal point position and deflection angle.

However, because of the size restriction of the pixel on SLM, there is a limited diffractive angle when it comes to reconstruction, which will influence the adjustable range of the pupils' interval. As is shown in fig 2 (b).

According to the grating function:

$$d \sin\theta = m\lambda$$

When m=1, the maximum diffractive angle of SLM can be written as:

$$\alpha_{max} = \sin^{-1}\frac{\lambda}{d}$$

Here d is the pixel pitch of SLM, and the maximum adjustable range of the interval can be written as:

$$\frac{d}{2} = f \cdot \tan(\alpha_{max})$$

Another dilemma of pupil duplication in Maxwellian display is the image shift when observer moves from one viewpoint to another one. This discontinuous condition also brings a challenge to its practicability.

As is shown in fig3 (a), the reason for discontinuity is the image from different viewpoints projecting on different districts of the retina. However, there are overlapping area between different projection areas. Once the images can coincide with each other in the overlapping area (as shown in fig3 (b)), the image shift effect can be eliminated.

Fig3 (c) and (d) shows the principle to align the images based on the central image. Here d is the pupil diameter, $f'$ is focal length of the lens, u is the distance from SLM to lens, $h_s$ is the half size of SLM, θ is the deflection angle of the upper image, h are the distance between the edge and the center point from the optical axis and $\theta_n$ are the angle from the converge point to the center of the image projected onto the retina.

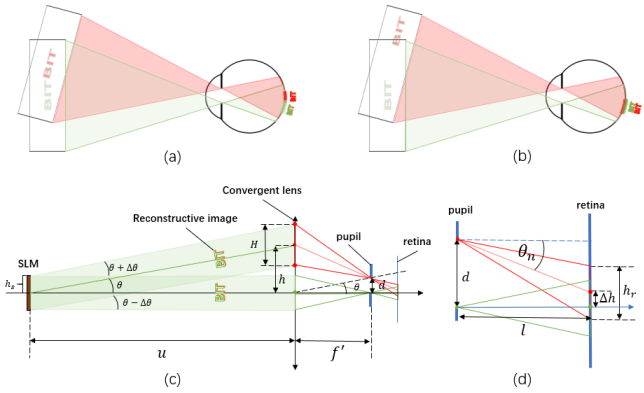

Fig. 3. The analysis of pre-compensation for continues eyebox. (a) Overlap analysis. (b) Schematic diagram to eliminate overlapping effects. (c)and(d) The principle of pre-compensation.

The key is to calculate the compensation Δh that makes the different images coincide on the retina. Δh can be written as:

$$\Delta h = d - l\tan\theta_n$$

And the remaining parameters can be obtained:

$$\theta = \arctan(\frac{h-d}{f'})$$

$$\Delta h = d - l\tan\theta_n$$

$$H = h_s + u \cdot (\tan(\theta+\Delta\theta) - \tan(\theta-\Delta\theta))$$

$$h = u\tan\theta$$

Supposing the resolution of the image in this view is W, the compensation N of the original image content is:

$$N = W \cdot \frac{\Delta h}{h_r}$$

Optical experiments were performed to verify the principle and the feasibility of the proposed method. Fig. 4 shows the schematic diagram and the picture of the experimental setup. In the experiments, λ = 532nm green laser diode (LD) was used as a light source for hologram display, and parallel beams were obtained via spatial filter elements and collimater lenses. The resolution of the SLM was 1,920×1,080 pixels, with a pixel pitch of 8μm and a refresh rate of 60Hz. CGHs were obtained using the point-based method, and a holographic image could be reconstructed in air when the SLM loaded with CGHs was illuminated by parallel beam. A 4f lens system and band-pass filter (BPF) were used to eliminate unwanted high order diffractions and zero-order diffraction.

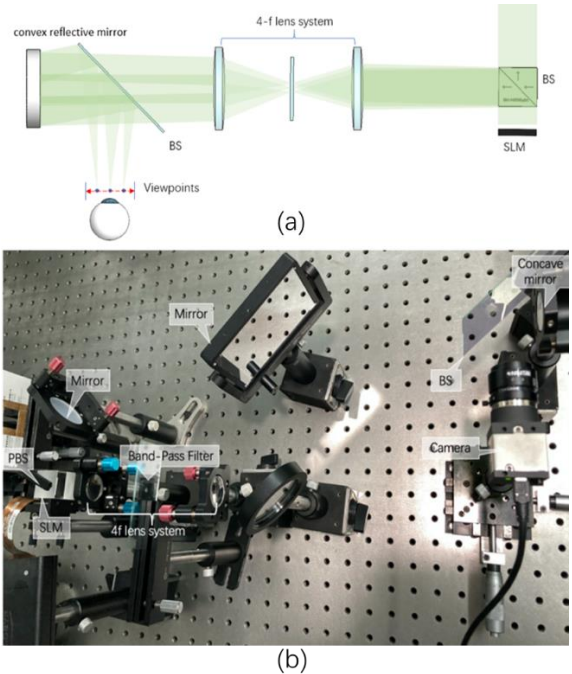

Fig. 4. Optical experiment system. (a) Schematic diagram of the system. (b) Actual experimental system.

Here in experiments, a see-through eyepiece consisting of a beam-splitter and a convex mirror was used to replace the convergent lens. The rays from SLM are transmitted through a half mirror and reflected by the convex mirror. The radius of the mirror was 75.8mm, with an effective focal length of 30.6mm, a diameter of 25.3mm, and the eye relief distance was 20mm from the eyepiece. The hologram virtual image was located 32cm from the convex reflective mirror, and it was reconstructed 75cm away from the SLM. The horizontal size of reconstruction image was 21mm, which was 226mm from the viewpoint. So, the horizontal FOV was 5.32°. Note that the reconstruction image would diverge from its position to convex mirror, and its size would be restricted by the diameter of the convex mirror. Optical experiment results of interval adjustment are shown in Fig. 5 (a). The interval can change continuously from 0mm to 3mm.

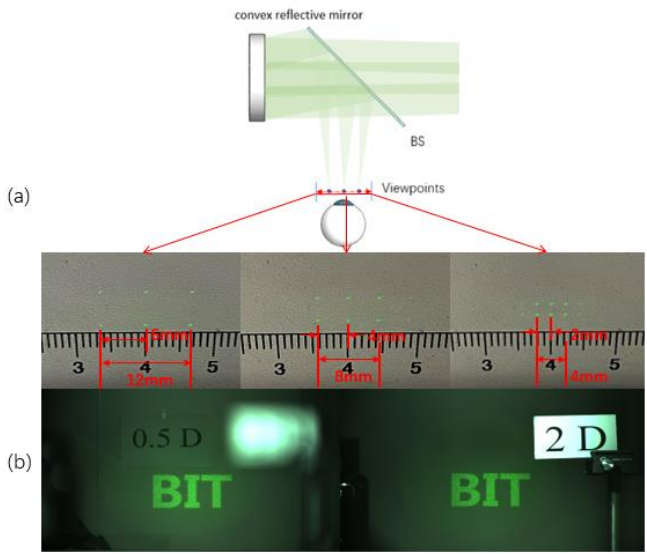

Fig. 5. Optical experiment results of interval adjustment (a) and Maxwellian display (b).

The replicated eyebox here consists of a constant 2*3 array, of which the maximum size was 9*6mm. In actual use, more convergent points can be added to the array to obtain a large enough eyebox according to the interval. Fig. 5 (b) shows the display effect of Maxwellian display through one convergent point of the array. The image in sight stay clear no matter when the camera focuses near or far.

Fig6 show the sights moving in the eyebox. The interval here was constant 1.5mm and the camera pupil was set as 1mm, 1.5mm and 2mm. The eyebox here consisted of a 1*3 array. When PD>interval, double image appeared because there were two spots located in pupil area at switching positions, as a2 and a4 show. Besides, blind area came from the positions where there were no spot in pupil as b2 and b4 show. c2 and c4 show the situation when PD matched with interval. They were at the positions where the image from last viewpoint began to disappear and the image from next viewpoint began to appear. However, overlap happened owing to the image shift among the different focal spots. Fig. 6 d show the effect after pre-compensation. As d2 and d4 show, overlap disappeared at the switching positions and the image kept continuous during the camera moving.

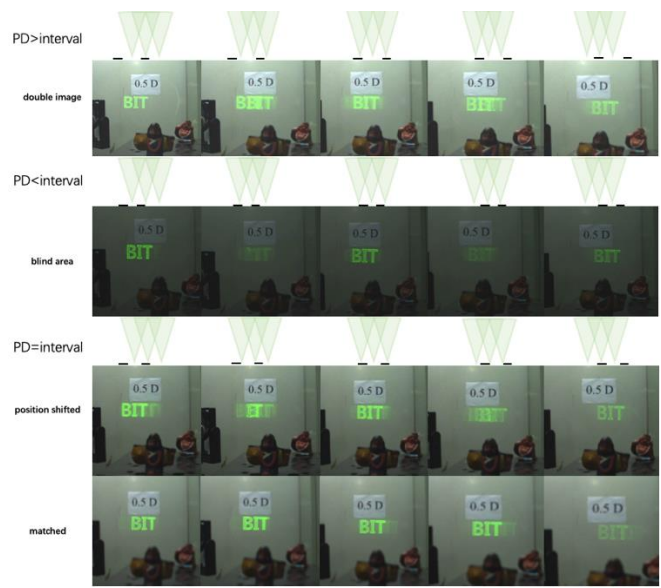

Fig. 6. Optical experiment results of interval match and pre-compensation.

Although the working principle of the proposed system was verified by the experiment, there still is large room for enhancement. It is worth noting that the brightness of images from different positions were not even because of the overlap. However, the probability of eyes being in overlapping positions is quite small especially when the pupils and interval match well. As for the brightness problem in other places, it comes from filtering, can be solved by proper filtering and hologram preprocessing in the future.

Apart from that, images also slightly tilted among the points because there are different convergent angles among the convergent points, which finally tilted the images. Nevertheless, the large depth of Maxwellian display makes it possible to eliminate or alleviate the tilt by aligning the image on the retina, which can be achieved by pre-processing on holograms in the future.

In conclusion, we proposed an optical see-through holographic Maxwellian near-eye display with adjustable eye-box replication. The proposed system uses holographic way to expand the eyebox and adjust the interval of the focal spots, which makes it possible to match any size of the human pupils and avoid double image or a blind area problem. The implemented experimental setup provides continuous images across the replicated eyeboxes and remove image shifts among the different viewpoints by aligning the image on the retina. We expect the proposed method can be used in the AR applications where always-focused information needs to be displayed on the real environment.